\documentstyle[11pt,fleqn]{article}
\newcommand{\cs}{\hspace{-16.5pt}}

\newcommand{\beq}{\begin{equation}}
\newcommand{\eeq}{\end{equation}}
\newcommand{\md}{{\rm d}}
\newcommand{\e}{{\rm e}}
\newcommand{\imu}{{\rm i}}
\newcommand{\eq}[1]{eq.~(\ref{#1})}
\newcommand{\bra}[1]{\langle #1 |}
\newcommand{\ket}[1]{| #1 \rangle}
\newcommand{\ds}[1]{{\rm d}\tilde{#1}}
\newcommand{\p}{{\scriptscriptstyle(+)}}
\newcommand{\m}{{\scriptscriptstyle(-)}}
\newcommand{\pa}{{\scriptscriptstyle\parallel}}
\newcommand{\pe}{{\scriptscriptstyle\perp}}

\newcommand{\T}{{\rm T}}
\def\ga{\mathrel{\mathchoice {\vcenter{\offinterlineskip\halign{\hfil
$\displaystyle##$\hfil\cr>\cr\noalign{\vskip1.5pt}\sim\cr}}}
{\vcenter{\offinterlineskip\halign{\hfil$\textstyle##$\hfil\cr>\cr
\noalign{\vskip1.0pt}\sim\cr}}}
{\vcenter{\offinterlineskip\halign{\hfil$\scriptstyle##$\hfil\cr>\cr
\noalign{\vskip0.5pt}\sim\cr}}}
{\vcenter{\offinterlineskip\halign{\hfil$\scriptscriptstyle##$\hfil
\cr>\cr\noalign{\vskip0.5pt}\sim\cr}}}}}

\title{\vspace*{-2cm}\hspace*{\fill} {\normalsize SI 94-15}\\[3cm]
{\huge\bf Collinear Asymptotic Dynamics for Massive Particles.\\
Reggeization and Eikonalization.}\thanks{This work has been supported
under the German-Polish agreement on bilateral scientific and
technological cooperation.}}
\author{ \\  \\ \\\cs
\bf{J.\,Boguszynski,\,H.\,D.\,Dahmen,\,R.\,Kretschmer,\,%
L.\,{\L}ukaszuk,\,L.\,Szymanowski}\\ \\\cs
Siegen University, Fachbereich Physik, Siegen, Germany\\\cs
and\\\cs
Soltan Institute for Nuclear Studies, Warsaw, Poland}
\date{December 1994}
\begin{document}
\maketitle
\vspace{5cm}

\begin{abstract}
\large{The dynamics of massive particles in the collinear high momentum
regime is investigated. Methods hitherto exploiting large time asymptotic
Hamiltonians in the Dirac picture for the treatment of infrared divergences
are adapted to the collinear asymptotic dynamics. The essential role of
time ordering of the dressing operators is brought out.}
\end{abstract}

\newpage

\section{Introduction}

For the treatment of infrared divergencies and the problem of the
Coulomb phase it has been known for a long time that perturbative
contributions up to infinite order have to be taken into account.
Kinoshita \cite{Kinoshita}, Lee and Nauenberg \cite{LeeNau} showed
that maintaining unitarity in the approximation is the key for
solving this problem. Khulish and Faddeev \cite{KulFad} pointed out
that both phenomena are dominated by the asymptotic dynamics of QED.
They separated out that part of the Hamiltonian which is essential for
large time scales.

The dynamics at large times is dominated by the interaction involving
the long wave length photons. In the Dirac picture the temporal
variation of their contribution to the interaction Hamiltonian is
small. Actually, for the energy of the photons going to zero the
corresponding interaction Hamiltonian density in momentum
space becomes stationary.

Similarly, photons collinear to electrons lead to parts in the
interaction density stationary in the limit of high momentum
\cite{Dahmen1}.

In the case of massive particles only in the high momentum limit the
stationary regime is sustained.

Adapting the methods of asymptotic dynamics \cite{Drell,Dahmen2,Dahmen3}
to the massive case we shall show that they lead to Reggeization
\cite{Eden,Collins,Mueller}. While in the treatment of infrared dynamics
time ordering of the dressing operator did not come into play, here it
will be of essential importance.

We shall also show that the other typical high energy approximation,
i.~e.\ eikonalization \cite{Fried}, can be obtained using our methods.

In this paper for developing the appropriate adaptation of our method
we study the $\varphi^3$ theory, since it is the simplest model for
which Reggeization has been demonstrated.

\section{Approximate Unitarity}
\setcounter{equation}{0}

Calculations of bremsstrahlung processes require in a unitary
approximation the inclusion of an infinity of orders in the coupling
constant to ensure a finite total cross section.

One way of reaching this aim is through an approximate
calculation of Dyson's time evolution operator $U(t,t_0)$
determined by the equation
\beq
\imu {\partial \over \partial t} U(t,t_0) = H_{\rm I}(t) U(t,t_0)
\eeq
with the well known formal solution
\beq
U(t,t_0) = \T \exp \left\{ -\imu \int \limits_{t_0}^t H_{\rm I}(t')
\, \md t' \right\} \quad,
\eeq
where $H_{\rm I}(t)$ is the interaction Hamiltonian in the Dirac
picture.

For definiteness we exemplify our method on the $\varphi^3$
model. The interaction Hamiltonian
\beq
H_{\rm I} = - {g \over 3!} \int {:} \varphi^3(x) {:} \, \md^3 x
\eeq
can be written as
\beq
H_{\rm I} = - {g \over 3!} \int \left\{ \varphi^{\m 3}(x) +
\varphi^{\p 3}(x) + 3 \varphi^{\m 2}(x) \varphi^\p(x) +
3 \varphi^\m(x) \varphi^{\p 2}(x) \right\} \md^3 x
\eeq
by splitting the field operator
$\varphi(x) = \varphi^\m(x) + \varphi^\p(x)$ into the
creation and annihilation parts.

As in the argument of Kulish and Faddeev \cite{KulFad} in the
case of bremsstrahlung and Coulomb phase we select those
contributions to the interaction that are most important at large
time scales. At first we neglect those terms of $H_{\rm I}$ that
create or annihilate three particles and get
\beq
\tilde{H}_{\rm I} = H_{\rm I}^\p + H_{\rm I}^\m \label{a11}
\eeq
with
\beq\label{g}
H_{\rm I}^\p = - {g \over 2} \int \varphi^{\m 2}(x)
\varphi^\p(x) \, \md^3 x \quad,\quad
H_{\rm I}^\m = - {g \over 2} \int \varphi^\m(x)
\varphi^{\p 2}(x) \, \md^3 x \quad.
\eeq
or in momentum space
\beq
H_{\rm I}^\p = - {g \over 4} \int {\ds{k} \, \ds{p} \over E(\vec{p} -
\vec{k})} \, a^\dagger(\vec {p} - \vec{k}) a^\dagger(\vec{k})
a(\vec{p}) \exp \left\{ \imu [ E( \vec{k}) + E( \vec{p} - \vec{k}) -
E(\vec{p})] t \right\}
\eeq
where
\beq
\ds{k}={\md^3 k \over (2\pi)^3 2k^0} \quad.
\eeq
Collinear asymptotic dynamics (for $p \gg m$, where $p = |\vec{p}|$)
will be incorporated by restricting the $\vec{k}$-integration to a region
$\Omega(\vec{p})$ in which $E(\vec{k}) + E(\vec{p} - \vec{k}) -
E(\vec{p})$ tends to zero for $p \to \infty$. A convenient choice for
$\Omega(\vec{p})$ is given by the requirement
\beq\label{a12}
k^\mu l_\mu \leq \sigma^2
\eeq
where $l^\mu$ is the momentum vector of the second created particle,
$l^\mu = (E(\vec{p} - \vec{k}), \vec{p} - \vec{k})$, and $\sigma$ is a
scale parameter. For $\sigma > m$ the region $\Omega(\vec{p})$ defined
in this way restricts $k^\pa$ and $\vec{k}^\pe$ to an ellipsoid
\beq
{\left( k^\pa - {p \over 2} \right)^2 \over a^2} +
{\vec{k}^{\pe 2} \over b^2} \le 1
\eeq
with the principal axes $a$ and $b$ given by
\beq
a^2  =  {p^2 \over 4} \left( {\sigma^2 - m^2 \over
\sigma^2 + m^2} \right) + {\sigma^2 - m^2 \over 2}
\quad \mbox{and} \quad
b^2  =  {\sigma^2 - m^2 \over 2} \quad.
\eeq
For $p \gg \sigma \gg m$ we have
\beq
a \approx {p \over 2} \quad.
\eeq
The calculation of $U(t,t_0)$ is greatly simplified if the
time-ordered exponential is suitably factorized. This is
facilitated by a time dependent similarity transformation.
For two operators $A(t)$, $B(t)$, not necessary Hermitean,
one has
\begin{eqnarray}
\lefteqn{\T \exp \left\{ (-\imu) \int \limits_{t_0}^t A(t')
\, \md t' +
(-\imu) \int \limits_{t_0}^t B(t') \, \md t' \right\} =}
& & \label{a} \\
& = & \T \exp \left\{ -\imu \int \limits_{t_0}^t A(t')
\, \md t' \right\}
\T \exp \left\{ -\imu \int \limits_{t_0}^t W_1^{-1}(t',t_0)
B(t') W_1(t',t_0) \, \md t'
\right\} \nonumber
\end{eqnarray}
with
\beq
W_1(t,t_0) = \T \exp \left\{ -\imu \int \limits_{t_0}^t A(t')
\, \md t'
\right\} \quad.
\eeq
The validity of the factorization of the T-product follows
from a factored ansatz $U= W_1 W$ for the time evolution
operator leading to the differential equation
\beq
\imu {\partial \over \partial t} W(t,t_0) = W_1^{-1}(t,t_0) B(t)
W_1(t,t_0) W(t,t_0)
\eeq
having again as solution a time-ordered exponential
\beq
W(t,t_0) = \T \exp \left\{ -\imu \int \limits_{t_0}^t
W_1^{-1}(t',t_0) B(t') W_1(t',t_0)
 \, \md t' \right\}
\quad.
\eeq
For further use we quote a formula already given by Dyson
\cite{Dyson}
\begin{eqnarray}
\lefteqn{W_1^{-1}(t',t_0) B(t') W_1(t',t_0) =} & & \\
& = & \sum_{n=0}^\infty \imu^n \int \limits_{t_0}^t \md t_n
\int \limits_{t_0}^{t_n}
\md t_{n-1} \cdots \int \limits_{t_0}^{t_2} \md t_1
[ A(t_1) , [ A(t_2) , \ldots
[ A(t_n) , B(t) ] \ldots ] \quad. \nonumber
\end{eqnarray}
In our context $A(t)$, $B(t)$ will be identified with
$H_{\rm I}^\p(t)$, $H_{\rm I}^\m(t)$, respectively. We will use
the truncated expression
\beq
W_1^{-1} B W_1 = B(t) + \imu \int \limits_{t_0}^t
[ A(t_1) , B(t) ] \, \md t_1
+ O(g^3) \quad.
\eeq
Again applying \eq{a} we can introduce
\beq
W_2(t,t_0) = \T \exp \left\{ -\imu \int \limits_{t_0}^t B(t')
\, \md t' \right\}
\eeq
and
\beq
W_3(t,t_0) = \T \exp \left\{ \int \limits_{t_0}^t \md t_2
\int \limits_{t_0}^{t_2} \md t_1 [ A(t_1) , B(t_2) ] \right\}
\eeq
(note that the time ordering in $W_3$ refers to $t_2$). In this
way we get an approximate expression for the time evolution
operator
\beq\label{b}
U(t,t_0) = W_1(t,t_0) W_2(t,t_0) W_3(t,t_0) \quad.
\eeq
It fulfills an approximate unitarity relation of the kind
\beq
U^\dagger(t,t_0) U(t,t_0) = \exp O(g^3) \quad.
\eeq
A similar approach has been successfully used in bremsstrahlung
calculations \cite{Dahmen1}. In what follows we are going to
show that the above approximation produces Reggeization and
eikonalization of two-body amplitudes for elastic scattering.

Let us give the explicit formula for  $W_3$ using
$A(t) = H_{\rm I}^\p(t)$, $B(t) = H_{\rm I}^\m(t)$:
\begin{eqnarray}
W_3(t,t_0) & = & \T \exp \left\{ {g^2 \over 4}
\int \limits_{x^0 = t_0}^t \md^4 x
\int \limits_{y^0 = t_0}^{x^0} \md^4 y \, \Bigl(
\varphi^{\m 2}(y) \varphi^{\p 2}(x) \Delta^+(y-x)
\right. \label{c} \\
& & \left. - 4 \varphi^\m(x) \varphi^\m(y)
\varphi^\p(x) \varphi^\p(y) \Delta^+(x-y)
\Bigr) \vphantom{\int \limits_{t_0}^t} \right\} \nonumber
\end{eqnarray}
with
\beq
\Delta^+(x-y) = [ \varphi^\p(x) , \varphi^\m(y) ] \quad.
\eeq
In the expression for $W_3$ we have neglected a part that
corresponds to renormalization corrections which will not be
considered here.

\section{Reggeization}
\setcounter{equation}{0}

In the case of soft electromagnetic bremsstrahlung the
asymptotic interaction Hamiltonian is linear in the photon field.
The photon creating part of this Hamiltonian corresponding to our
$H_{\rm I}^\p(t)$ commutes with itself at different times,
since the electromagnetic current can be well approximated by the
classical one. Thus, the time-ordering of the exponential is
superfluous.

In contradistinction, $H_{\rm I}^\p(t)$, \eq{g}, is a product
of $\varphi^{\m 2}$ and $\varphi^\p$ and therefore does not
commute with itself at different times, so that the time-ordering
in $W_1$ remains essential. A similar comment holds true for
$W_2$.

Let us start with a suitable formula for the elastic amplitude
${\cal T}_{\rm if}(p_1 + p_2 \to p_3 + p_4)$:
\beq\label{k}
{\cal T}_{\rm if} = T + T^{\rm cr} \quad,
\eeq
where
\beq\label{h}
T = - \imu \int \md^4 x \, \md^4 y \, \e^{- \imu p_2 y +
\imu p_4 x} \Theta(x^0 - y^0)
\bra{\vec{p}_3, {\rm out}}
j_{\rm H}(x) j_{\rm H}(y) \ket{\vec{p}_1, {\rm in}}
\eeq
and
\beq
T^{\rm cr} = T(p_2 \leftrightarrow - p_4) \quad.
\eeq
In \eq{h} the Heisenberg current is denoted by the
index ``H''.

The matrix element $T$ can be written as
\begin{eqnarray}
T & = & - \imu \int \md^4 x \, \md^4 y \,
\e^{- \imu p_2 y + \imu p_4 x} \Theta(x^0 - y^0) \label{i1} \\
& & \times
\bra{\vec{p}_3}U^\dagger(x^0, -\infty) j(x) U(x^0,y^0)
j(y) U(y^0, -\infty)
\ket{\vec{p}_1} \quad. \nonumber
\end{eqnarray}
In what follows we shall consider small transfers
$t = (p_1 - p_3)^2$ at high energies. It will be convenient to
work in the laboratory frame of the particle with momentum $p_2$;
$p_1$, $p_3$ are chosen to be large momenta. Detailed calculations
will be made in the interaction picture and we expect that the leading
contribution to the reggeized amplitude is real and comes from the
quasi-stable multiparticle configurations. Therefore it is reasonable
to replace
\beq
U(x^0,y^0) \to {\bf 1} \quad.
\eeq
Inserting a complete set of states between $j(x)$ and
$j(y)$ yields
\beq
T = (2\pi)^4 \delta(p_1 + p_2 - p_3 - p_4) {\cal M}_2 \quad,
\eeq
with
\beq
{\cal M}_2 = (2\pi)^3 \sum_{n=1}^\infty {1 \over n!} V_n
\label{a15} \quad,
\eeq
where
\beq\label{i}
V_n = \int \ds{k}_1 \cdots \ds{k}_n \,
{\delta(\vec{p}_1 + \vec{p}_2 - \sum_{i=1}^n \vec{k}_i)
\over p_1^{\ 0} + p_2^{\ 0} - \sum_{i=1}^n k_i^{\ 0} +
\imu \varepsilon} \, G^\ast_n(\vec{p}_3) G_n(\vec{p}_1)
\eeq
and
\beq
G_n(\vec{p}) = \bra{\vec{k}_1, \ldots, \vec{k}_n} j(0)
U(0,-\infty) \ket{\vec{p}} \quad. \label{a13}
\eeq
As we are working in the laboratory frame of the particle with momentum
$p_2$, the current $j(0)$ is connected with the interaction of
this slow particle. Since we expect that $U(0,-\infty)$ alone accounts
for the production of energetic particles, we replace
\beq
j(0) = g \left( \varphi^\m(0) \varphi^\p(0) + {\varphi^{\m 2}(0) +
\varphi^{\p 2}(0) \over 2} \right)
\eeq
by
\beq
j^{\rm s}(0) = g \varphi^\m(0)\varphi^\p(0)
\eeq
in \eq{a13}.

Next we substitute the approximation for $U(0,-\infty)$, cf.\ \eq{b}.
The operators $W_2$ and $W_3$ act trivially on the one-particle
state $\ket{\vec{p}}$:
\beq
W_2 W_3 \ket{\vec{p}} = \ket{\vec{p}} \quad.
\eeq
So we are left with
\beq\label{m}
G_n(\vec{p}) = g \bra{\vec{k}_1, \ldots, \vec{k}_n} \left(
\varphi^\m(0) \varphi^\p(0) \right) \T \exp
\left\{ -\imu \int \limits_{-\infty}^0 \md t \, H_{\rm I}^\p(t)
\right\} \ket{\vec{p}} \quad.
\eeq
Let us remark that we can limit ourselves to the term of order
$(n-1)$ in the time ordered expansion in \eq{m}. Lower orders lead to
disconnected graphs which should not contribute to the full
amplitude in the asymptotic collinear dynamics considered here. The
terms higher than $(n-1)$ necessarily lead to vertex and/or mass
corrections neglected throughout this paper.

Therefore
\begin{eqnarray}
G_n(\vec{p}) & = & g (-\imu )^{n-1} \int \limits_{-\infty}^0 \md t_1
\cdots \md t_{n-1} \, \Theta(t_1, \ldots , t_{n-1}) \label{n} \\
& & \times \bra{0} a(\vec{k}_1) \cdots a(\vec{k}_n) \varphi^\m(0)
\varphi^\p(0) H_{\rm I}^\p(t_1) \cdots H_{\rm I}^\p(t_{n-1})
\ket{\vec{p}} \quad, \nonumber
\end{eqnarray}
where the symbol $\Theta(t_1, \ldots, t_{n-1})$ denotes
\beq
\Theta(t_1, \ldots , t_{n-1}) = \Theta(t_1 - t_2) \cdots
\Theta(t_{n-2} - t_{n-1}) \quad.
\eeq
It is useful to rewrite $G_n$ as
\beq\label{o}
G_n(\vec{p}) = \sum_{r=1}^n g^n_r(\vec{p})
\eeq
where
\begin{eqnarray}
g^n_r & = & g (-\imu)^{n-1} \int \limits_{-\infty}^0 \md t_1 \cdots
\md t_{n-1} \, \Theta(t_1, \ldots , t_{n-1}) \label{p} \\
& & \times \bra{0} a(\vec{k}_1) \cdots a(\vec{k}_{r-1})
a(\vec{k}_{r+1}) \cdots a(\vec{k}_n) \varphi^\p(0)
H_{\rm I}^\p(t_1) \cdots H_{\rm I}^\p(t_{n-1}) \ket{\vec{p}}
\quad. \nonumber
\end{eqnarray}
In evaluating the matrix elements in \eq{p} we have to commute
the $a(\vec{k}_i)$ with $H_{\rm I}(t_j)$. Thus we shall encounter
the following commutators of $a(\vec{k})$:
\begin{eqnarray}
D(t,\vec{k}) & = & -\imu [ a(\vec{k}) , H_{\rm I}^\p(t) ] =
\imu g \int_{x^0=t} \md^3 x \,
\varphi^\m(x) \varphi^\p(x) \e^{\imu k x} \quad, \\
d(t,\vec{k},\vec{k}\,') & = & [ a(\vec{k}) , D(t,\vec{k}\,') ] =
\imu g \int_{x^0=t} \md^3 x \, \varphi^\p(x)
\e^{\imu (k + k') x} \quad, \\ {}
[ a(\vec{k}\,'') , d(t,\vec{k},\vec{k}\,') ] & = & 0 \quad.
\end{eqnarray}
The operator $d$ is linear in $\varphi^\p$. Therefore we should
consider the commutator of $d$ with $H_{\rm I}^\p$ again. Since $d$
depends on the sum of two intermediate momenta it leads to vertex
and mass corrections which---as already stated above---are beyond
the scope of our work. As a consequence, $d$ can be treated as
a commuting quantity and it is easy to see that for this reason it
does not contribute at all. In this way we obtain
\begin{eqnarray}
\lefteqn{(-\imu)^{n-1} \bra{0} a(\vec{k}_1) \cdots a(\vec{k}_{r-1})
a(\vec{k}_{r+1}) \cdots a(\vec{k}_n) \varphi^\p(0) H_{\rm I}(t_1)
\cdots H_{\rm I}(t_{n-1}) \ket{\vec{p}} =} & & \\
& & = \sum_{\pi_r}
\bra{0} \varphi^\p(0) D(t_1, \vec{k}_{\pi(1)}) \cdots
D(t_{n-1},\vec{k}_{\pi(n)}) \ket{\vec{p}} \quad, \nonumber
\end{eqnarray}
where the summation runs over all permutations of the numbers
$1, 2, \ldots, r-1, r+1, \ldots, n$.

Applying consecutively
\beq
D(t,\vec{k}) \ket{\vec{p}} = \imu g {\e^{\imu (E(\vec{p} -
\vec{k}) + k^0 - p^0) t} \over 2 E(\vec{p} - \vec{k})}
\ket{\vec{p} - \vec{k}} \quad,
\eeq
and integrating, we arrive at
\begin{eqnarray}
\lefteqn{g^n_r(p) =} & & \label{r} \\
& = & g^n \sum_{\pi_r}
{1 \over 2 E(\vec{p} - \vec{k}_{\pi(n)})} \times
{1 \over E(\vec{p} - \vec{k}_{\pi(n)}) + k_{\pi(n)}^{\  0} - p^0 }
\nonumber \\
& & \times {1 \over 2 E(\vec{p} - \vec{k}_{\pi(n)} -
\vec{k}_{\pi(n-1)})} \times
{1 \over E(\vec{p} - \vec{k}_{\pi(n)} - \vec{k}_{\pi(n-1)}) +
k_{\pi(n)}^{\  0} + k_{\pi(n-1)}^{\ 0} - p^0 } \nonumber \\
& & \quad\quad \vdots \nonumber \\
& & \times {1 \over 2 E(\vec{p} - \sum_{1, i \neq r}^n
\vec{k}_{\pi(i)})} \times
{1 \over E(\vec{p} - \sum_{1, i \neq r}^n \vec{k}_{\pi(i)}) +
\sum_{1, i \neq r}^n k_{\pi(i)}^{\ 0} - p^0 } \quad. \nonumber
\end{eqnarray}
Having this form of $g_r^n$, we now look for the kinematical
region, that gives a dominant contribution. Note that for massive
particles the denominators in $g^n_r$ never vanish.

We shall calculate the dominant contributions in the leading
logarithmic approximation, so we consider fixed transverse
momenta and
\begin{eqnarray}
\left| k^\pa_{\pi(s)} \right| & \gg & \sqrt{\vec{k}_{\pi(s)}^{\pe 2}
+ m^2} \\
\left| q^\pa_{r,s} \right| & \gg & \sqrt{\vec{q}_{r,s}^{\pe 2} + m^2}
\end{eqnarray}
for
\beq
s = 1, 2, \ldots, r-1, r+1, \ldots, n
\eeq
and with
\beq
q_{r,s} = p - \sum_{i=s, i \ne r}^n k_{\pi(i)} \quad.
\eeq
Then one can show that the terms
\beq
 {1 \over 2 E(\vec{q}_{r,s}) [E(\vec{q}_{r,s}) - q^0_{r,s}]}
\eeq
in \eq{r} are different from zero for
\beq
k^\pa_{\pi(s)} \gg q^\pa_{r,s} \label{a14}
\eeq
or
\beq
k^\pa_{\pi(s)} = O(q^\pa_{r,s}) \quad.
\eeq
The second possibility does not contribute in leading
logarithmic order, so we are left with the conditions \eq{a14}.
These, together with momentum conservation (see \eq{i})
\beq
k^\pa_r = q^\pa_{r,1}
\eeq
lead to strong ordering of the $k$'s
\beq\label{j}
p^\pa \ga k^\pa_{\pi(n)} \gg k^\pa_{\pi(n-1)} \gg \ldots
\gg k^\pa_{\pi(r+1)} \gg k^\pa_{\pi(r-1)} \gg \ldots \gg
k^\pa_{\pi(1)} \gg k^\pa_r \gg m \quad.
\eeq
It means that in the expression \eq{i} (with $\vec{p}^\pe_1 = 0$)
only the same orderings in both $G_n$ enter, i.~e.
\begin{eqnarray}
V_n & = & g^{2n} \int \ds{k}_1 \cdots \ds{k}_n \,
{\delta(\vec{p}_1 - \sum_1^n \vec{k}_s) \over
p_1^{\ 0} + p_2^{\ 0} - \sum_{i=1}^n k_i^{\ 0}} \,
\sum_{r=1}^n \sum_{\pi_r} f^n_r(\pi) \\
& & \times {1 \over m^2 + (\vec{p}_3^\pe - \vec{k}_{\pi(n)}^\pe)^2}
\times
{1 \over m^2 + \vec{k}_{\pi(n)}^{\pe 2}} \cdots \nonumber \\
& & \times {1 \over m^2 + (\vec{p}_3^\pe - \sum_{1, s \neq r}^n
\vec{k}_{\pi(s)}^\pe)^2} \times
{1 \over m^2 + (\sum_{1, s \neq r}^n \vec{k}_{\pi(s)}^\pe)^2} \nonumber
\end{eqnarray}
where $f^n_r(\pi)$ is unity in the region defined by \eq{j} and
zero otherwise.

Integrating over transverse momenta we arrive at
\beq
V_n = {g^2 \over 2 (2 \pi)^3} \left( g^2 K(\vec{p}_3^{\pe 2})
\right)^{n-1} \sum_{r=1}^n \sum_{\pi_r} I_r^n(\pi)
\eeq
where
\beq
I_r^n(\pi) = \int {\md k_1^\pa \cdots \md k_n^\pa \over
k_1^{\ 0} \cdots k_n^{\ 0}} \times
{f^n_r(\pi) \delta(p_1^\pa - \sum_1^n k_i^\pa) \over
p_1^{\ 0} + p_2^{\ 0} - \sum_1^n k_i^{\ 0}}
\eeq
and
\beq
K = {1 \over 2 (2 \pi)^3} \int {\md^2 k^\pe \over
(m^2 + \vec{k}^{\pe 2})(m^2 + (\vec{p}_3^\pe - \vec{k}^\pe)^2)}
\quad.
\eeq
The longitudinal integration yields the same result for each
$I^n_r$:
\beq
I^n_r(\pi) = {2 \over s} {(\ln s)^{n-1} \over (n-1)!}
\left( 1 + O \left( {1 \over \ln s} \right) \right)
\eeq
where $s \approx 2 m p_1^\pa$, so that
\beq
V_n = n! {g^2 \over (2 \pi)^3} {(g^2 K \ln s)^{n-1}
\over (n-1)!} {1 \over s}
\quad.
\eeq
Then using \eq{a15} results in
\beq
{\cal M}_2 = {g^2 \over s} \sum_{n=1}^\infty
{(g^2 K \ln s)^{n-1} \over (n-1)!} = g^2 s^{-1 + g^2 K} \quad,
\eeq
i.~e.\ we have obtained Regge behaviour \cite{Collins}.

\section{Eikonalization}
\setcounter{equation}{0}

In this chapter we apply the factorization formula \eq{b}
to derive the familiar eikonal form of the scattering amplitude.

We start with
\beq
\bra{\vec{p}_3, \vec{p}_4} S \ket{\vec{p}_1, \vec{p}_2}
\quad,\quad S = U(\infty,-\infty) =
W_1(\infty,-\infty) W_2(\infty,-\infty) W_3(\infty,-\infty)
\;\;.
\eeq
In this matrix element the part $W_1 W_2$ can be replaced by
unity; other terms of $W_1 W_2$ either give a one-particle
$s$-channel contribution that can be neglected, or vanish
identically, because $W_3$ conserves the particle number.
The first term in the exponent of $W_3$, \eq{c}, is also
non-leading for large c.\ m.\ s.\ energy.

In this way we arrive at
\beq\label{d}
\bra{\vec{p}_3, \vec{p}_4} S \ket{\vec{p}_1, \vec{p}_2} =
\bra{\vec{p}_3, \vec{p}_4} \T \exp \left\{
\int \limits_{-\infty}^\infty r(t) \, \md t + O(g^3) \right\}
\ket{\vec{p}_1, \vec{p}_2} \quad,
\eeq
where
\beq
r(t) = - g^2 \int \limits_{-\infty}^t \md y^0 \int \limits_{x^0=t}
\md^3 x \, \md^3 y \, \varphi^\m(x) \varphi^\m(y) \varphi^\p(x)
\varphi^\p(y) \Delta^+(x-y) \quad.
\eeq
In fact, for large energies the time ordering in \eq{d} is
irrelevant. To see this, consider any two-particle state
$\ket{\vec{k}_1, \vec{k}_2}$ in the center of mass system
of $k_1$ and $k_2$. Then we have
\begin{eqnarray}
r(t) r(t') \ket{\vec{k}_1, -\vec{k}_1}
& = & - { g^4 \over 4} \int \ds{q} \, \ds{q}' \,
a^\dagger(\vec{k}_1 + \vec{q} + \vec{q}\,')
a^\dagger(-\vec{k}_1 - \vec{q} - \vec{q}\,') \ket{0} \label{e} \\
& & \times {\e^{-2 \imu (k_1^{\ 0} - E(\vec{k}_1 + \vec{q})) t}
\over
(E(\vec{k}_1 + \vec{q}))^2 [k_1^{\ 0} - q^0 - E(\vec{k}_1 +
\vec{q}) + \imu \varepsilon]} \nonumber \\
& & \times {\e^{-2 \imu (E(\vec{k}_1 + \vec{q}) - E(\vec{k}_1 +
\vec{q} + \vec{q}\,')) t'} \over
(E(\vec{k}_1 + \vec{q} + \vec{q}\,'))^2 [E(\vec{k}_1 + \vec{q}) -
q^{\prime 0} - E(\vec{k}_1 + \vec{q} + \vec{q}\,') +
\imu \varepsilon]} \; . \nonumber
\end{eqnarray}
In the limit $k_1^{\ 0} \to \infty$, \eq{e} reduces to
\begin{eqnarray}
r(t) r(t') \ket{\vec{k}_1, -\vec{k}_1}
& \stackrel{k_1^{\ 0} \to \infty}{=} &
- {g^4 \over 4 (k_1^{\ 0})^4} \int \ds{q} \, \ds{q}' \,
a^\dagger(\vec{k}_1 + \vec{q} + \vec{q}\,')
a^\dagger(-\vec{k}_1 - \vec{q} - \vec{q}\,') \ket{0} \nonumber \\
& & \times {\e^{2 \imu q^\pa t} \over q^0 + q^\pa -
\imu \varepsilon} \times
{\e^{2 \imu q^{\prime \pa} t'} \over q^{\prime 0} + q^{\prime \pa} -
\imu \varepsilon} \quad,
\end{eqnarray}
from which it is evident that
\beq
[ r(t) , r(t') ] \ket{\vec{k}_1, -\vec{k}_1}
\stackrel{k_1^{\ 0} \to \infty}{=} 0 \quad.
\eeq
As a consequence, \eq{d} can be written as
\beq
\bra{\vec{p}_3, \vec{p}_4} S \ket{\vec{p}_1, \vec{p}_2} =
\bra{\vec{p}_3, \vec{p}_4} \exp \left\{
\int \limits_{-\infty}^\infty
r(t) \, \md t \right\} \ket{\vec{p}_1, \vec{p}_2}
\eeq
with
\beq\label{f}
\int \limits_{-\infty}^\infty r(t) \, \md t
\, \ket{\vec{p}_1, \vec{p}_2} =
\imu {g^2 \over 8 (2\pi)^2 (p_1^{\ 0})^2}
\int {\md^2 q^\pe \over \vec{q}\,^{\pe 2} + m^2}
a^\dagger(\vec{p}_1 + \vec{q}\,^\pe) a^\dagger(\vec{p}_2 -
\vec{q}\,^\pe) \ket{0} \quad.
\eeq
{}From \eq{f} it is straightforward to derive the well-known
eikonal formula
\beq
{\cal M}_{\rm f{}i} = 2 \imu s \int \md^2 b \,
\e^{\imu \vec{p}_3^\pe \vec{b}}
\left[ \exp \left( {\imu g^2 \over 4 \pi s}
K_0(m |\vec{b}|) \right) - 1 \right] \quad.
\eeq

\section{Concluding remarks}

We have shown how the collinear asymptotic dynamics described
by the Hamiltonian \eq{a11} restricted to the kinematical region
$\Omega$, \eq{a12}, leads to both Reggeization and eikonalization
of two-particle scattering. While Reggeization is dominated by
three-particle collinear dynamics, eikonalization corresponds to
that kinematical region of the asymptotic Hamiltonian where only
two particles are collinear. In contradistinction to infrared dynamics
\cite{KulFad,Dahmen2}, the time ordering of the dressing operator
is essential for three-particle collinear dynamics which leads to
Reggeization.

\section*{Acknowledgement}
Three of us (HDD, RK, LS) would like to acknowledge
Prof.\ Lev Lipatov for discussions.

\end{document}